\documentclass[12pt,preprint]{aastex}

\usepackage{epstopdf}

\shorttitle{Photoevaporation and spatial variation of grain sizes in Orion 114-426}
\shortauthors{Miotello et al.}

\begin{document}

\title{Evidence of photoevaporation and spatial variation of grain sizes \\
in the Orion 114-426 protoplanetary disk
}

\author{Anna Miotello\altaffilmark{a} and Massimo Robberto}
\affil{Space Telescope Science Institute, Baltimore, MD 21218, USA}
\author{Marco A. C. Potenza}
\affil{Dipartimento di Fisica, Universit\`a degli Studi di Milano and INFM, Via Celoria, 16, Milano I-20133, Italy}
\author{L. Ricci}
\affil{Caltech, Cahill Center for Astronomy and Astrophysics,  1200 East California Blvd,
Pasadena CA 91125, USA}

\altaffiltext{a}{Universit\`{a} degli Studi di Milano, Italy}

\begin{abstract}
Deep HST broad-band images taken with ACS and WFPC2 of the giant ($\sim 1000$~AU diameter) dark silhouette proplyd 114-426 in the Orion Nebula  show that this system is tilted, asymmetric, warped and photoevaporated. The exquisite angular resolution of ACS allows us to map the distribution of dust grains at the northern translucent edge of the disk, dominated by the photoevaporative flow.
Using the Mie theory for standard circumstellar disk grains, we find evidence for a spatial gradient in grain size. The typical dust radius, $\simeq 0.2-0.7~\mu$m (less than what reported by previous studies) becomes smaller as the distance from the disk center increases, consistent with the expectations for the dynamic of dust entrained in a gaseous photoevaporative wind. 
Our analysis of the disk morphology and location within the nebula indicates that this system is photoevaporated by the diffuse radiation field of the Orion Nebula, while being shielded from the radiation coming directly from the central Trapezium stars. We  estimate the mass-loss rate from the disk surface and the time-scale for total disk dissipation, which turns out to be of the order of $10^4$yr. Such a short time, of the order of 1/100 of the cluster age, indicates that this system is seen on the verge of destruction. This is compatible with the exceptional nature of the disk, namely its combination of huge size and low mass. Finally, we briefly discuss the viability of possible mechanisms that may lead to the peculiar morphology of this system: external UV flux, binary star and past close encounter.
\end{abstract}


%________________________________________________________________________________________________________________

\section{Introduction}

%One of the most critical aspects of star and planet formation is the lifetime of young circumstellar disks. 
Early \emph{Hubble Space Telescope} (HST)  images of circumstellar disks in the Orion Nebula Cluster have provided the first direct evidence that protoplanetary disks in rich young stellar clusters may be photo-evaporated by UV radiation of nearby massive stars  \citep{ODellWen94}. The typical signature of photo-evaporation is 
a bright ionized cusp pointing to the brightest stars of the cluster, $\theta^1$Ori-C in particular. The cusps always surround a central low-mass star, that may occasionally be encircled by a compact dark disk seen in silhouette against the nebular background, or against the cusp itself.
The most recent atlas of protoplanetary disks (``proplyds'') in the Orion Nebula \citep{Ricci} counts 178 of these comet-shaped sources. The same atlas also counts other 36 disks which do not show evidence of photoevaporation, being seen as pure dark silhouettes either in absorption against the nebular background (28 sources) or as dark lanes at the midplane of some diffuse bipolar emission (8 sources). 
Disk photo-evaporation naturally raises the issue of disk survival in the harsh environment of young massive clusters  \citep[see e.g.][]{ODell+08}. The frequency and spatial distribution of disks with and without evidence of photo-evaporation within the same cluster provides an excellent  opportunity to investigate how environmental factors may affect the path to planet formation. \\

%The majority of the 40 {\bf sono davvero 40 o e' cifra tonda?} disks catalogued by \cite{O'Dell2} are photoevaporated, showing a
%However, the list also contains other 15 sources seen purely in silhouette against the background light of the Orion Nebula. Their typically larger sizes, from 25 to 500 AU, match the  expected sizes for protoplanetary disks surrounding young low-mass stars \citep{Johnstone}.

The basic physical process at the origin of disk photo-evaporation in massive clusters has been clarified by  \citet{Johnstone}. The FUV (6 eV $\le h\nu \le$ 13.6 eV; 2000\AA$\ge\lambda\ge$ 912\AA) radiation reaches nearly undisturbed the disk surface, creating a photodissociation region (PDR) with temperature $T\sim100-1000$\,K. Depending on the radial distance from the ionizing stars, the heated disk surface may reach thermal velocity higher than the escape velocity. The resulting neutral outflow, largely composed by molecular $H_2$ gas dragging a non negligible amount of small dust grains \citep{Owen} soon encounters the EUV  ($h \nu \ge$ 13.6 eV; $\lambda \le$ 912\AA) radiation coming directly from the ionizing stars, creating a ionization front at $T\simeq10^4$~K, i.e. the bright ionized cusp. This model can explain both the general morphology of photoevaporated disks as well as more peculiar features, like the presence of [OI] emission from photodissociation of OH radicals  produced by $H_2$ heating and dissociation observed in one of the most prominent objects, 182-413 \footnote{In this paper we shall adopt the coordinate-based naming convention of \citet{ODell+08}.} \citep{StorzerHollenbach99, Bally+00}.  

It is generally assumed that disk photo-evaporation in rich clusters, as revealed by the presence of photo-ionized cusps, can occur only if the disk is directly exposed to the EUV flux from the massive stars, consistently with the fact that the majority of photo-evaporated disks in the Trapezium cluster lies in the immediate surroundings of $\theta^1$Ori-C. Dark silhouette disks, lacking any evidence of photo-ionization, are believed to lie within the foreground Veil of the Orion Nebula, shielded by EUV radiation \citep{McCaughreanOdell96}. However, dark silhouette disks can also be photo-evaporated by FUV radiation \citep{Adams+04} although the products of photo-dissociation in these systems have never been seen \citep{ODell+08}. This because the parameter controlling photo-evaporation is the intensity of the FUV flux, which can be expressed in units of \emph{Habing Flux} (the estimate average flux in the local interstellar medium), defined as $G_0 = 1.6 \times 10^{-3} $erg cm$^{-2}$ s$^{-1}$. 
In the inner core of the Trapezium, the FUV flux is elevated, reaching $G = 10^5 G_0$ at the Orion Bar \citep{Robberto+02}, about  110\arcsec or 0.22~pc south of $\theta^1$Ori-C \citep[assuming a distance $d = 414$~pc][]{Menten+07}\footnote{while \citet{Menten+07} provide the most precise estimate, \citet{ODellHenney08} consider a distance $d=430\pm20$~pc to be more accurate. Our estimates  do not depend on the exact value of the assumed distance.}. 
Sources in the outskirt of the region, shielded from the EUV photons emitted e.g. by $\theta^1$Ori-C, may still receive scattered EUV and, especially, FUV radiation from the nebular environment.

The most spectacular case of dark silhouette in Orion is the well know 114-426 disk \citep{McCaughrean98}. With an apparent diameter of over $2\arcsec$, which makes it visible also from the ground, this disk spans about $\sim 1000$ AU, 10 times the size of our Solar System or of other typical photo-ionized disks in Orion. It is seen nearly edge-on, the presence of the central star being noticeable through scattered polar emission, and is considered to be in a relatively quiescent, non photo-evaporated phase \citep{Throop+01}.  \citet{McCaughrean98} estimate for the central star an initrinisic brighness $K\sim9.5^m$, corresponding to a 1.5M$_\odot$ star at the distance and age of the Orion cluster, but the uncertainites are large. Using HST/NICMOS high-resolution near-infrared images of 114-426, \cite{McCaughrean98} find that the major axis of the disk in the Pa$\alpha$ at 1.87 $\mu$m line is $\sim$20\,$\%$ smaller than at 0.6\,$\mu$m. From this, they deduce a minimum disk mass of 10\,M$_\oplus$, and possibly $\geq 5 \times 10^{-4}$\,M$_{\odot}$. On the other hand,  114-426 has not been detected by the Submillimeter Array survey  \citep{Mann} at either 880\,$\mu$m or 1330\,$\mu$m. This non-detection places an upper mass limit of $ 1.2 \times 10^{-2} M_{\odot}$ (0.5 - 12 $M_{Jup}$), confining the disk mass into a range of two orders of magnitude.

This object has attracted interest because the variation of the disk size with wavelength has allowed to probe the properties of the disk grains, exploiting the wavelength dependence of the extinction of the bright nebular background through the translucent disk edges \citep{Throop+01}. 
If the grain sizes were much smaller than the observed wavelengths, as in the Rayleigh scattering limit, shorter wavelengths would be scattered much more efficiently than  longer ones: in this case the disk would appear smaller at long wavelengths. Viceversa, if  the grain sizes are of the same order of the wavelength of observation, the disk size is expected to remain nearly constant with wavelength (``gray'' scatter).  Using observations of  114-426 in the H$\alpha$ (0.656 $\mu$m; HST/WFPC2), Pa$\alpha$ (1.87 $\mu$m; HST/NICMOS1) and Br$\alpha$ (4.05 $\mu$m; Keck/NIRSPEC), \citet{Shuping} found evidence for chromatic extinction at the northern translucent edge of the disk, deriving a typical grains diameter $\gtrsim 1.9 \mu$m, but not $\gg$ 4 $\mu$m.  This represents one of the first evidences for grain growth in protoplanetary disks,  the initial step in the process of planet formation \citep{Natta+07}.

In this paper we use the most recent HST/ACS images of the 114-426, together with a HST/WFPC2 image taken in the U-band, to clarify the disk structure and its physical status. Contrary to what is commonly assumed, we show that the disk appears strongly affected by the action of the environment and is photoevaporated. Its unusually large size and morphology (warped and asymmetric with respect to the central star) may result from photoevaporation. The translucent northern edge of the disk appears originated by the outflow of neutral material rather than the truncated edge of an undisturbed disk. We map the variation of extinction in the translucent region, showing how the typical grain size variation is compatible with an evaporative flow. This allows us to estimate the mass-loss rate from the disk surface and the disk lifetime. We also show that the disk is enshrouded by a large scale ($\sim0.03$~pc)  cocoon of dust  with a bright ridge facing the side opposite to the cluster center. This suggests that the disk photoevaporation is driven by the diffuse radiation field rather than by direct irradiation from the massive cluster stars. 

In Section~\ref{sec:Observations} we briefly present our data; in Section~\ref{sec:Morphology} we illustrate the geometry of the source, moving from the inner structure to the most extended features and the nebular surroundings; in Section~\ref{sec:Dust}  we analyze the grain size composition and distribution in the photoevaporated flow; in Section~\ref{sec:Discussion} we derive the main physical parameters of the mass loss and discuss the possible heating sources. Our conclusions are presented in Section~\ref{sec:Conclusions}.

%________________________________________________________________________________________________________________

\section{Observations}\label{sec:Observations}

HST/ACS-WFC and HST/WFPC2 images of 114-426 have been collected as part of the HST/Treasury Program on the Orion Nebula (GO-10246, PI M. Robberto). In this paper we use the data taken in six broad-band filters: F336W with WFPC2, and F435W, F555W, F658N, F775W, F850LP with ACS (Fig.~\ref{fig1}). The passbands are approximately equivalent to the standard $U$, $B$, $V$, $H\alpha$, $I$ and $z$ photometric system, respectively, and for simplicity we will equivalently use both naming conventions.

The instrument configuration, observation and data reduction strategy are presented elsewhere \citep{Robberto+12}. Here we remind that the ACS images have been drizzled to a scale 50 mas/pixels, corresponding to the nominal  scale of the instrument, whereas the WFPC2 images have been coadded preserving the original 10\, mas/pixel scale of the PC chips. %After careful examination, a shift of 1 pixel has been applied to the F658N image to match the position of the source in the other ACS filters. 
The FWPC2 image has been resampled and rotated to align it to the ACS images. A column of bad pixels in the WFPC2 image survived image processing and is visible in the F336W image. However, it's location and extent does not affect our analysis.

\section{Morphology of 114-426}\label{sec:Morphology}
\subsection{Disk Structure}

\begin{figure}
%\epsscale{2.5}
\centering
%\plotone{immagini_definitive/all_filters.pdf}
\includegraphics*[viewport=1 236 430 520, width=0.5\textwidth, page=1]{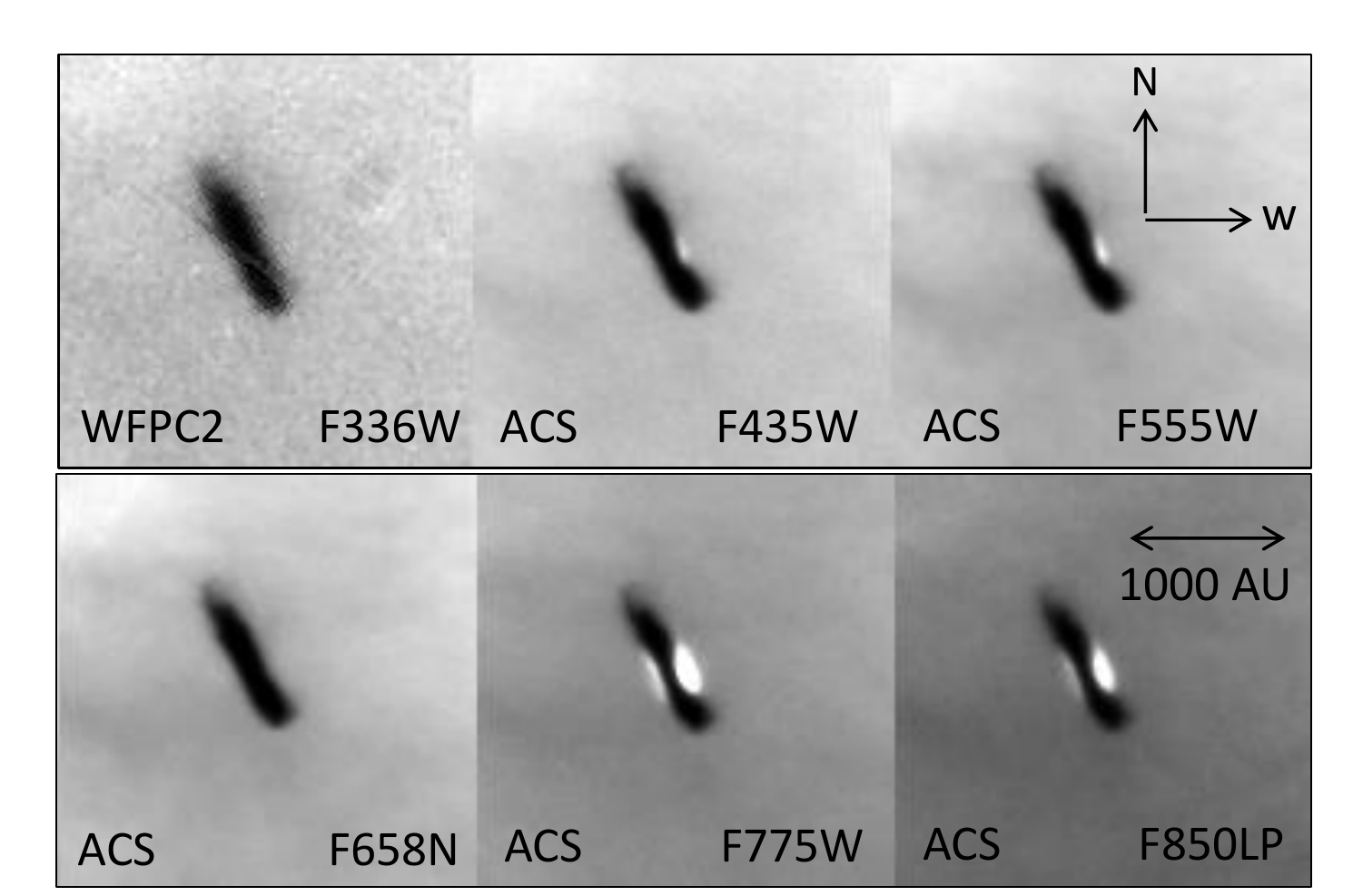}
%Viewport: The first two coordinates are the (x,y) coordinates in pixels of the lower left corner of the portion of the image file you want to include. The second two coordinates are the upper right values of (x,y)
\caption{HST images of the protoplanetary disk 114-426 in our six photometric filters, equivalent to the standard $U$, $B$, $V$, $H\alpha$, $I$ and $z$ passbands. }
\label{fig1}
\end{figure}

In Fig.~\ref{fig1} we present our six images ordered  in order of increasing wavelength. The scattered light from the central star,
leaking through the polar lobes of the disk, becomes more prominent with wavelength. The western bright lobe is always brighter than the eastern one, indicating that the disk, seen nearly edge-on, is tilted with the western side facing us. The morphology also shows clear signs of multiple asymmetries. First, the lobes are not symmetric with respect to the disk's polar axis, being displaced from the center and closer to the southern edge. Second, the lobes are not on the same vertical axis: a line joining the centers of the two lobes would clearly appear tilted with respect to the disk axis, with  an inclination of $\sim 45^{\circ}$. 

\begin{figure*}
\epsscale{.8}
\centering
%\plotone{immagini_definitive/3warped.eps}
\includegraphics*[viewport=1 1 240 300, width=0.3\textwidth, page=1]{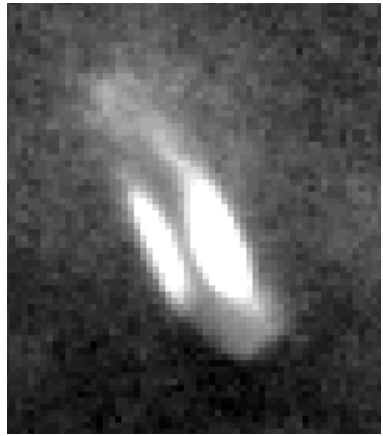}
\caption{Difference of images of 114-426 taken in F658N and F850LP filters, showing the warped geometry of the system. }
\label{fig2}
\end{figure*}

To further investigate the structure of this system, we have subtracted from the F850LP image the F658N image, properly scaled, to enhance the image contrast by exploiting the cromatic variation of dust opacity (Fig.~\ref{fig2}). The disk appears warped, with an ``equatorial''  dark lane tracing in silhouette the asymmetry of the bright lobes. 
The southern edge of the disk is  remarkably sharp and thick, at a projected distance of $\sim370$~AU  from the mid-point between the two lobes, which may be assumed as the position of the central sta . We will take $r_d=400$~AU as the nominal radius of the disk. 

\begin{figure*}
\epsscale{1.2}
\centering
%\plotone{immagini_definitive/spur_def.eps}
\includegraphics*[viewport=1 150 500 420, width=0.7\textwidth, page=1]{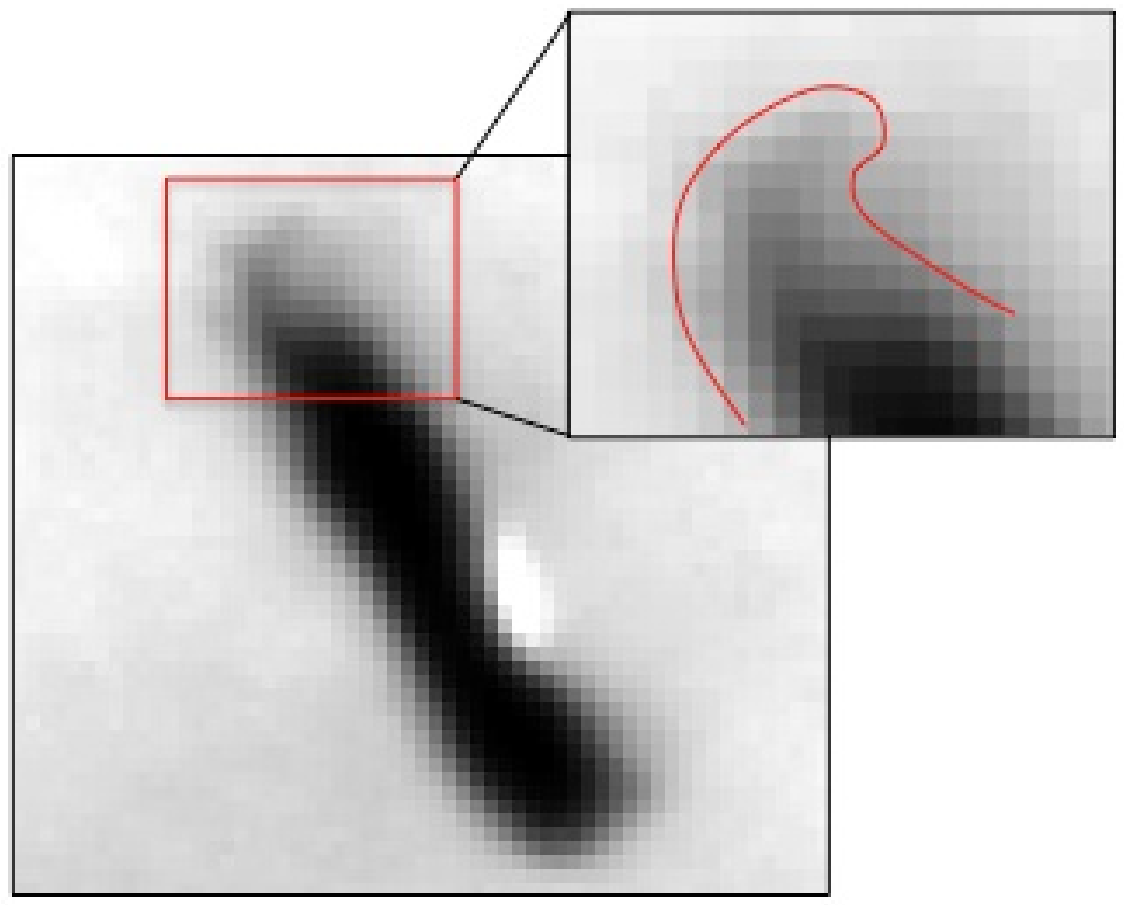}
\caption{Image of 114-426 in F555W filter. The upper edge of the disk presents a "spur-shaped" protrusion with a projected dimension of $\sim 250$ AU.}
\label{fig3}
\end{figure*}

\begin{figure*}
\epsscale{.8}
\centering
%\plotone{immagini_definitive/modelli.pdf}
\includegraphics*[viewport=1 310 730 530, width=0.6\textwidth, page=1]{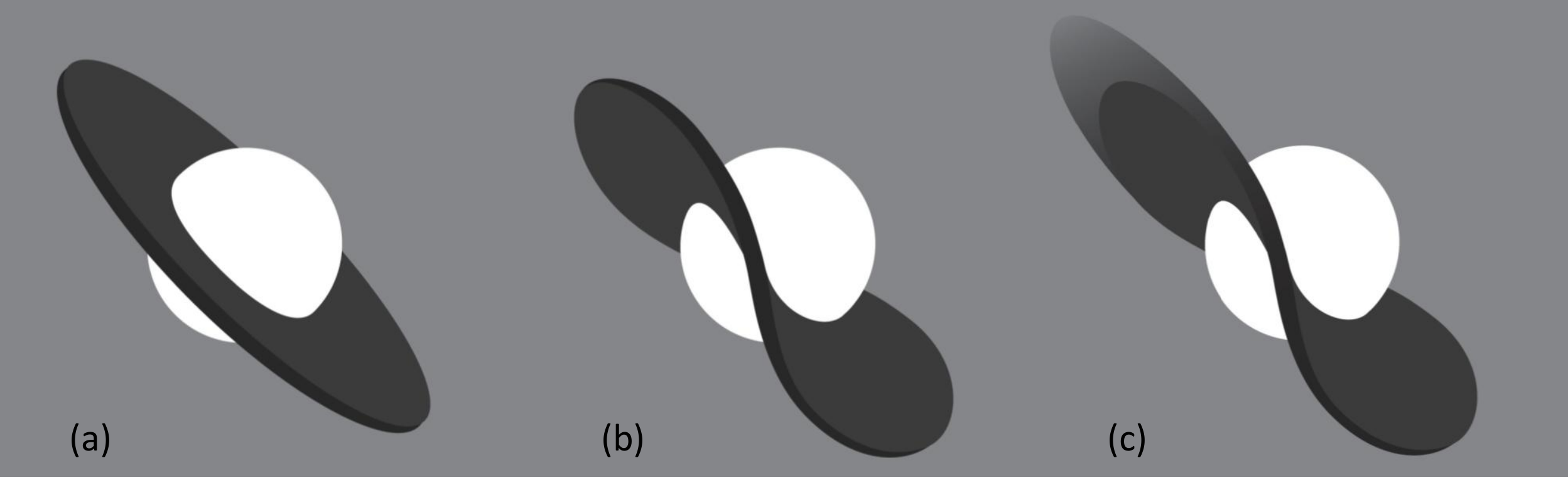}
\caption{Three graphic models of the tilted (a), warped (b) and photoevaporated shape of the disk.}
\label{fig2.a}
\end{figure*}

In what concerns the northern edge of the disk, by stretching the gray levels of our F555W image we can resolve it in a  spur-shaped structure (Fig.~\ref{fig3}) protruding beyond the darkest part of the disk. If we regard the spur as an outer appendix to the main disk, thus limiting the disk extension at the northern side, the symmetry of the warped disk with respect to the mid-point between the two lobes (i.e. the position of the central star) is nicely recovered.  We believe that the spur most probably represents an outflow of photoevaporated material. It corresponds to the "translucent" part of the disk where \citet{Throop+01} and \citet{Shuping}  have measured the dust grain size. In Figure~\ref{fig2.a} se shows a cartoon of the system summarizing the main features: tilt, asymmetry and warp. 

\subsection{Dark filaments}

\begin{figure*}
\epsscale{1}
\plotone{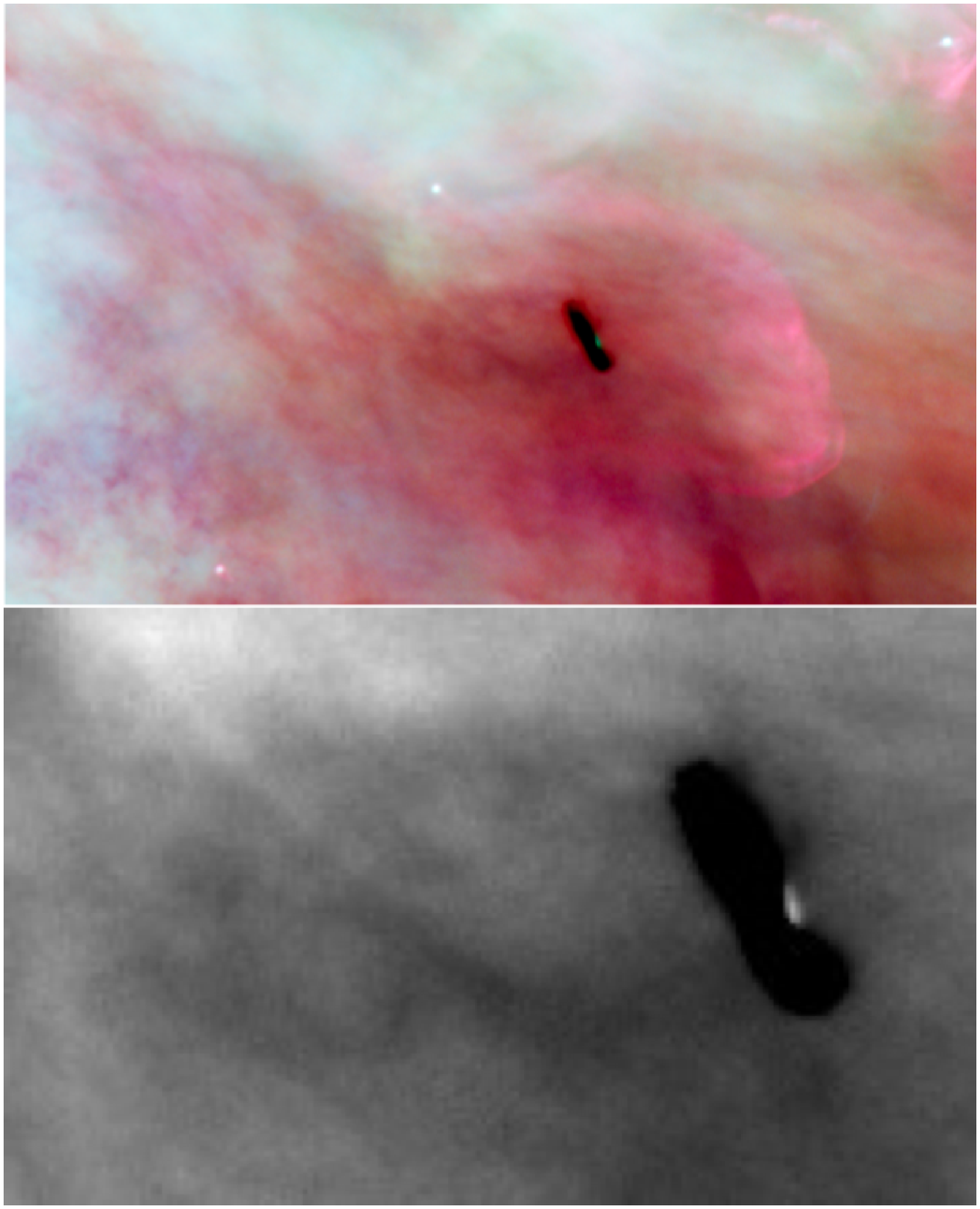}
\caption{Top: Color composite image of 114-426 in the F435W (blue), F555W (green) and F658N (H$\alpha$, red) filters; Bottom: difference of the F555W-F435W images, zoomed-in on the target. Both images show  evidence of dark filamentary "tails" of photoeaporated dust escaping from the disk. The disk size, about 2\arcsec diameter, provides the images scale.}
\label{fig4}
\end{figure*}

\begin{figure*}
\epsscale{1.0}
\plotone{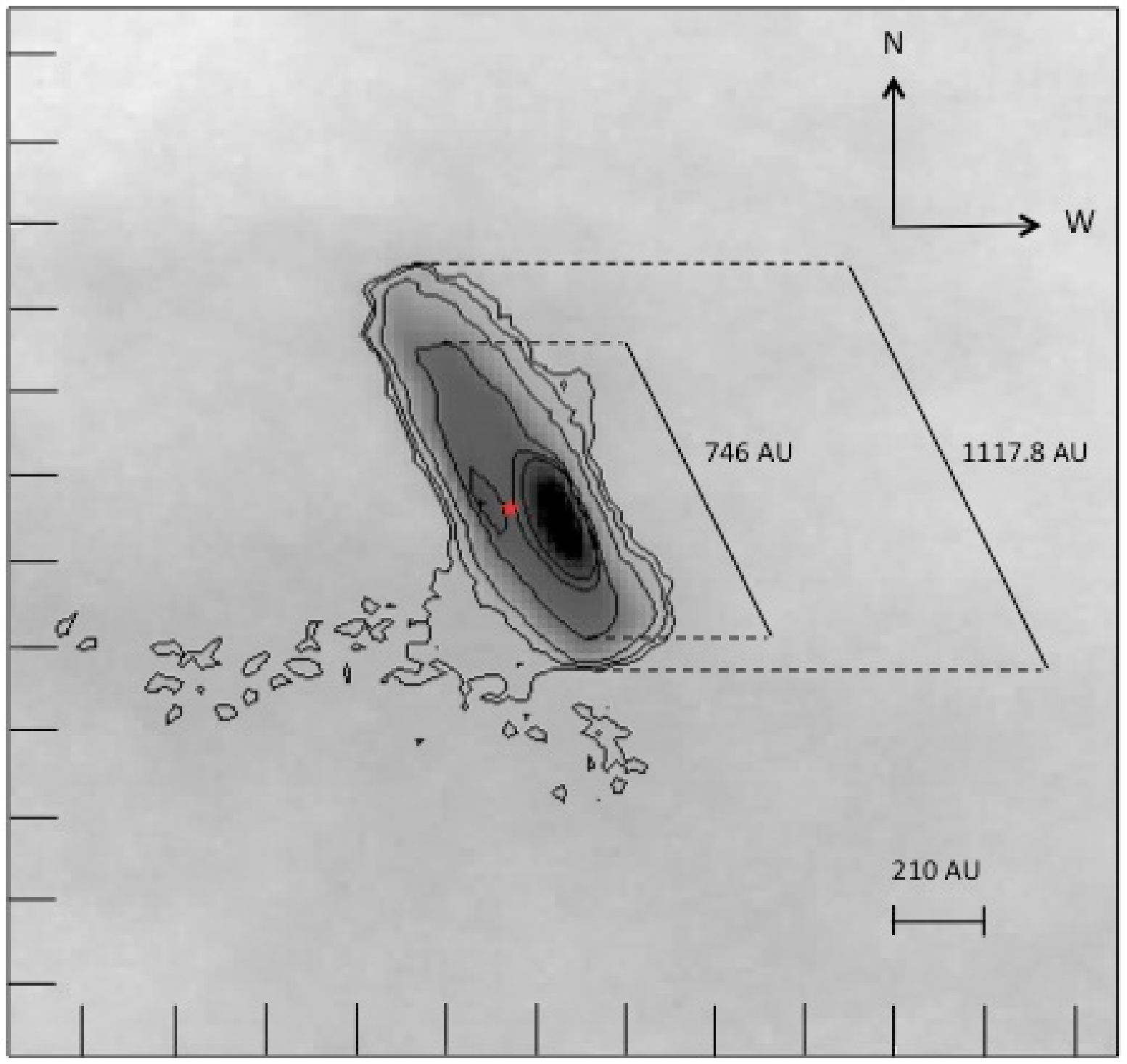}
\caption{Countour plot of 114-426 in the F658N (H$\alpha$) filter, showing the antisymmetry of the dark filaments with respect to the location of the bright lobes.}
\label{fig8}
\end{figure*}

Observing on a larger scale the H$\alpha$ image of the source, one can detect further signspots of ouflow of neutral material. Fig.~\ref{fig4} shows both a color composite image and a contranst enhanced V-B image of the target, stretched to enhance the contrast against the nebular background. Both images reveals a remarkable dark "tail" on the eastern side, characterized in the innermost region by twisted filaments that can be traced up to distances of several disk diameters.  The darker filament departs from the southern part of the disk, just to the south of the eastern lobe, and can be directly revealed by a contour plot of the disk intensity levels (Fig.~\ref{fig8}). The same figure also shows a minor protusion on the north-western side of the disk, just to the north of the bright lobe.  Notice that these two dark protrusions appear simmetric with respect to the disk center and antisymmetric with respect to the bright lobes.

\subsection{114-426 in its Environment\label{sec:foot}}

\begin{figure*}
\epsscale{1.2}
\plotone{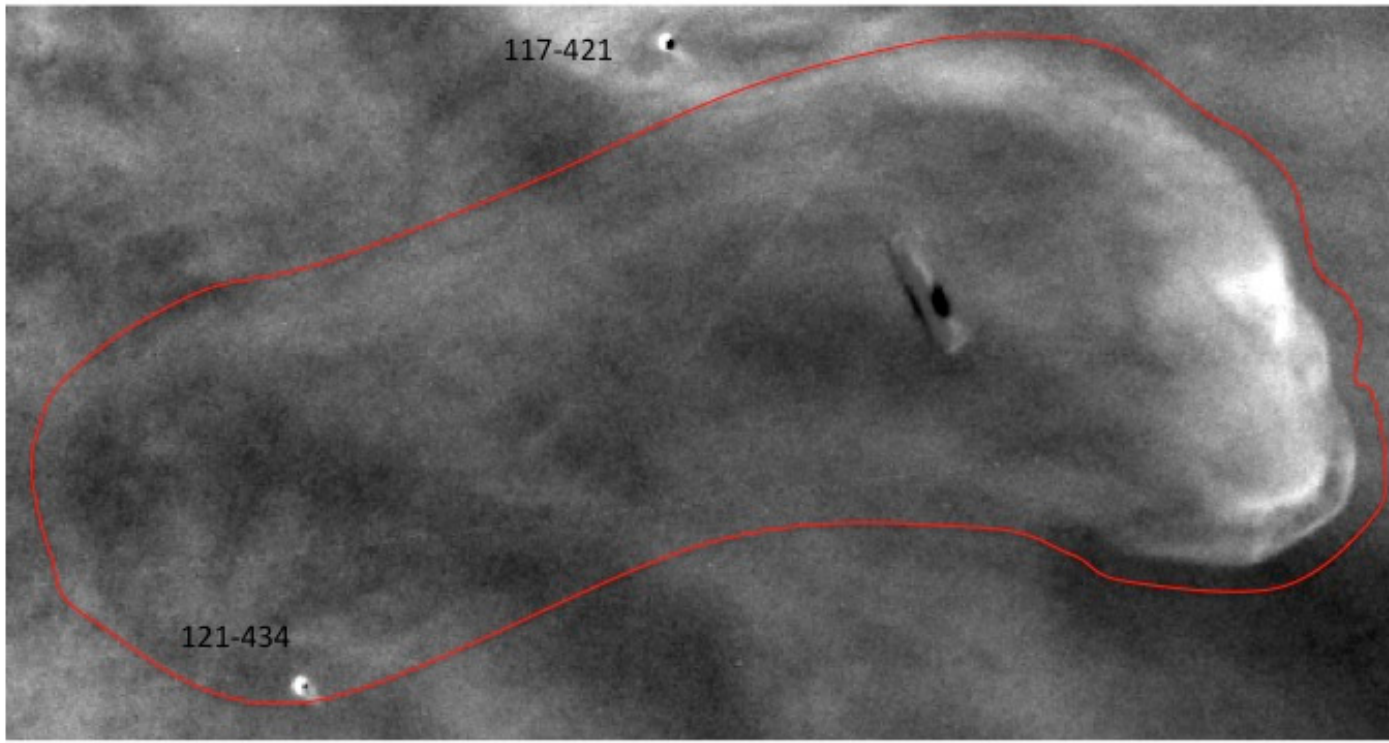}
\caption{The ``foot-like'' structure encircling 114-426, as revealed in the H$\alpha$ image. The size of this field is approximately $30\arcsec\times15\arcsec$.}
\label{fig5}
\end{figure*}

Figure~\ref{fig5} shows our source with its immediate surroundings in a contrast enhanced $H\alpha$-V image.  An extended filamentary shell emitting in $H\alpha$ line encircles the western side of the disk. The shell radius has a projected distance of about 2.5 disk diameters, $\sim 0.01$ pc. The filaments can be traced, at much lower brightness levels, to the eastern side of the disk, up to a projected distance of 0.03~pc. The tenuous bright rim delineates a characteristic ``foot-like" structure which seems to represent a region of enhanced density  externally illuminated by the diffuse radiation field of the nebula.

Remarkably, the two closest sources immediately to the east of 114-426,  117-421 and 121-434 \citep{Ricci}, are resolved in our ACS images in two compact photo-ionized proplyds, pointing in the direction of $\theta^1$Ori-C, lying about 100\arcsec\ to the north-east ($\simeq$0.2 pc projected distance). 
The small  size of the two dark disks, unresolved in our ACS images and in any case less than $\sim50$~AU radius (bounded by the ionized cups), is in striking contrast with the huge dimensions of our target, %seems to indicate that theire exposure to the EUV flux is not recent \citep{Johnstone}.\\
about 20 times larger in size and without any sign of direct influence from  $\theta^1$Ori-C.  \\

In conclusion, there are various elements of circumstantial evidence all pointing in the directions of 114-426 being photoevaporated, and we uphold that this is indeed the case. Before envisioning a possible scenario for this unique object, it is appropriate to reconsider, on the basis of our data, the physical properties of the dust grains. This is the subject of the next section.

%\begin{figure*}
%\epsscale{1.3}
%\plotone{immagini_definitive/confronto.eps}
%\caption{ }
%\label{fig9}
%\end{figure*}

\section{Dust particles sizing}\label{sec:Dust}
As shown by \cite{Throop+01} and  \citet{Shuping}, it is possible to investigate the properties of dust particles of 114-426 via the wavelength dependence of the extinction of background light in partially-opaque areas of the disk. Following the same type of analysis, 
%we obtain for a representative pixel the extinction coefficients presented in Figure~\ref{fig7} as a functions of $1/\lambda$. A comparison with the standard interstellar extinction curve immediately shows that the typical grain size in the disk is larger than that of ISM particles, as previously found by \citep{Shuping}. Our data, however, do not seem to confirm the high opacity reported by these authors at longer wavelenghts. {\bf unire Figure 7 e 9 come dice Luca; delle 3 curve di Figura 7, lasciare solo quella centrale, che dovrebbe rifersi al caso tipico, con R=3.1. Ovviamente, chiarire a quale parte del disco si riferisce; caso mai usare i dati di figura 11 per il pixel [46,88]} \\
we are now able to exploit 
the spatial resolution of our ACS images to map for the first time the particle size in different positions of the partially-opaque part of the disk.

We use to this purpose the classic Mie theory, a fully analytical model which derives, for the special case of spherical particles, an exact series expansion of the scattered field \citep[see for example][]{Hulst}. 
Three parameters are needed for computing the cross section of dust: the radius $a$, the real part of the refractive index $Re(n)$, and the imaginary part of the refractive index $Im(n)$, the latter two quantities being dependent on the wavelength of the incoming light. Both the absorption ($C_{abs}$) and the scattering ($C_{sca}$) cross sections are thus determined, and therefore the extinction cross section given by $C_{ext} = C_{abs} + C_{sca}$. Generally speaking, the dominant role is played by the radius $a$, while the refractive index components have a relatively minor effect. This fundamental property allows performing a measurement of the grain size from the shape of the cross section curve as a function of wavelength. 
The refractive index of the grains can be estimated given the dust grain composition, which is typically considered to be a compound of different chemical species \citep[e.g.][]{Pollack+04, Semenov+03}. The Bruggeman mixing rule  \citep{Bruggeman1935} provides a generic form that expresses the dielectric function of the  heterogeneous particle, $\epsilon_{0,eff}$, as a combination of the dielectric functions of the single species. For a $N$-components mixture it is:

\begin{equation}
\sum_{i=1}^{N}f_i\frac{\epsilon_{0,i}-\epsilon_{0,eff}}{\epsilon_{0,i}+2\epsilon_{0,eff}} = 0
\label{eq:Bruggeman}
\end{equation}

where $\epsilon_{0,i}$ and $f_i$ are the dielectric function and the volume density of the $i-th$ species ($\sum_{i=1}^Nf_i=1$), respectively. 
For the dust particles of circumstellar disk we consider porous composite grains made of astronomical silicates ($f=0.07$), carbonaceous materials ($f=0.21$), water ice ($f=0.42$) and vacuum ($f=0.30$). This is an approximated version of the \citet{Pollack+04} model \citep[see][]{Ricci2}. The dielectric function is derived from the complex refraction index, $\epsilon_i=n_i^2$. We have assumed for the Silicates the refraction index vs. the wavelength tabulated by  \cite{WeingartnerDraine2001} for the Silicates, by \cite{Zubko1996} for the amorphous Carbonacious material and by  \cite{Warren1984} for the Water Ice. For vacuum, $\epsilon_{0,vac}=1$. By solving the implicit Equation~\ref{eq:Bruggeman} for $\epsilon_{eff}$ we derive  
%, $n_i$, can then be estimated through the {\it Clausius--Mossotti--Lorentz--Lorenz} formula: 
%\begin{equation}
%4\pi\alpha_iN_i = \frac{3(n_i^2-1)}{n_i^2+2}
%\end{equation}
%where $N_i$[atoms/volume]$=\rho_i {N_A \over A_i}$ is the number density of the atoms,  while $\rho_i$, $N_A$ and $A_i$ are the mass density, Avogadro number and mass number, respectively. The constant $\alpha$ is the atomic polarizability. For the sake of simplicity we assume for both silicates and carbonates $Re(n) \sim 1.7{\bf Anna/Marco, cosa scrivo qui?} $, $\rho \sim 2.5$  $g/{cm^3}$, $A \sim 60$; for water ice $Re(n) = 1.33$, $\rho = 0.917$, $A = 18$. The polarizabilities can then be obtained, and we derive 
for the real part of the refractive index $Re(n)\simeq1.3$ and an imaginary part, $Im(n)\simeq 0.03$. 
%The imaginary has a limited impact on the cross sections, the main role being played by the grain radius. For the sake of simplicity one can adopt for our heterogeneous grains $Im(n) \simeq 0.01$, as it is commonly assumed in the literature (see for example \cite{Draine}).  
Having derived the effective refraction index (or dielectric function), we derive the absorption and scattering cross sections, and therefore the extinction, from the Mie theory.

The shape of the dust grains deserves some discussion. Real dust grains are not spherical but likely present complex internal structures due to the aggregation of smaller interstellar grains. The most recent studies   \citep{Min+06,Borghese} indicate that the scattering properties of particles of different shape and sub--structure are not dramatically different from those derived using the Mie theory for spheres. %Even if care has to be taken to properly determine the particle radius {\it a} , 
Assuming that the dust particles are perfect spheres does not introduce uncertainties larger than a 30\%, the size returned by the classic Mie theory being typically underestimated. 

Also the size distribution plays a role in determining the extinction. In this paper we make the simplifying assumption of treating the dust grains as if they were monodisperse, just one radius, in each pixel.  In a separate paper the issue of the size distribution will be treated more in detail. 

To compare theoretical predictions with observations, we exploit the extinction efficiency factor, $Q_{ext}$, namely the cross section divided by the geometrical cross section $G=\pi a^2$. This is a parameter commonly used in scattering problems, and can easily be obtained on the basis of the Mie expansion series as a function of the reduced size parameter $x = k a$, where $k = 2 \pi / \lambda$. The extinction curve can then be obtained at any $\lambda$ for a given size $a$, and then fitted to the data.% as detailed below. 
The extinction data have been evaluated as the ratio of the average flux ({\it counts/s}) of the nebular background, $F_{\lambda, back}$, and the flux ({\it counts/s}) in a single pixel of the translucent zone\footnote{The $\lambda$ index labels the broad-band filter.}, $F_{\lambda, pix}$. In particular $F_{\lambda, back}$ has been evaluated in an elliptic ring, all around the disk, close enough to be rapresentative of the effective nebular background of 114-426.
%, but far enough to avoid the counts coming from the central star.
More precisely we define $Q_{ext}$, given by the observational data, as

\begin{equation}
Q_{ext} = {ln({ F_{\lambda, back} \over F_{\lambda, pix}})\over G\Sigma}.
\end{equation}

\begin{figure*}
\epsscale{1.15}
\plotone{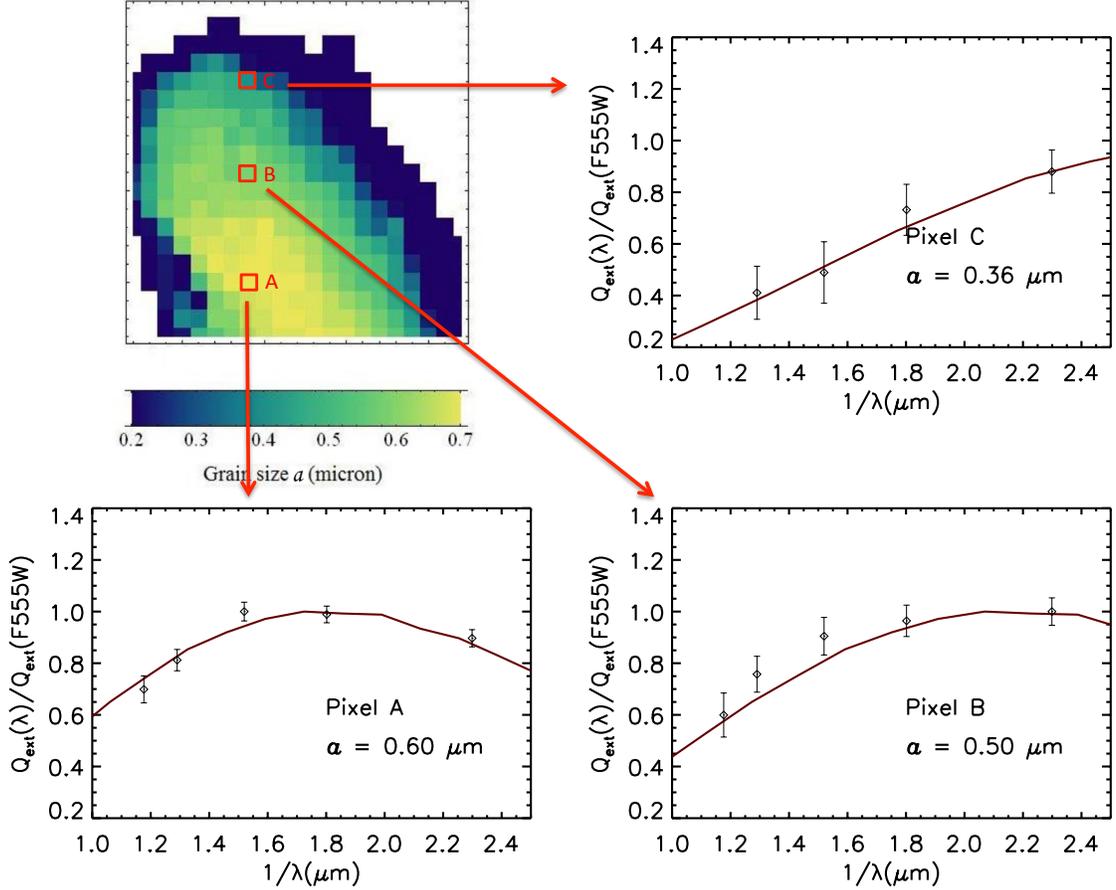}
%\plottwo{Pixel_Atrim.pdf}{Pixel_Etrim.pdf}
\caption{Size particles distribution in the photoevaporated area at the NE edge of 114-426. Different colors represent distinct mean radius $a$.  The three plots show the measured  absorption efficiency and corresponding extinction curves for the labeled pixels. The image is pixelated according to the ACS/WFC image scale, 1~pixel=50~mas=21~AU at d=414~pc. The coordinates of Pixel A are RA(2000.0)=5:35:11.345; DEC(2000.0)=-5:24:25.70, from the ACS images of the Orion Nebula obtained by the HST Treasury Program (file {\rm orion\_hst\_acs\_Strip0l\_f658n\_v1.fits} on http://archive.stsci.edu/prepds/orion/).}
\label{NewMap}
\end{figure*}

The surface density $\Sigma$ disappears when $Q_{ext}$ is normalized to some reference wavelength, for example the V-band (F555W). The plots in Fig.~\ref{NewMap} 
%{\bf qui si puo' citare la precedente figura 9, dove ora pero' ci concentriamo sulla linea "spessa" che fitta i nostri dati, cioe' la attuale figura 11} 
show the data for three representative pixels  and the corresponding best fit curves, providing radii equal to 0.60~$\mu$m, 0.50~$\mu$m and  0.36~$\mu$m. The attenuation in the V-band (F555W) varies from  $F_{V, back}/F_{V, pix}=4.0$ (pixel A) down to $F_{V, back}/F_{V, pix}=1.17$ (pixel C).  We estimated the error bars accounting for both the Poisson statistics (on $F_{\lambda, back}$ and $F_{\lambda, pix}$) and non uniformity of the background (for $F_{\lambda, back}$).
Similar plots have been obtained for every pixel in the translucent zone,  allowing us to map the distribution of dust grains size, also shown in Fig.~\ref{NewMap}.   From the outmost pixels (blue) to the innermost pixels (light yellow), the radii increase from  0.2 $\mu$m to 0.6 $\mu$m. The inner parts of the disk cannot be studied with this approach since the optical thickness increases and the scattered light from the lobes dominates. 

The absolute values of the radius $a$ we have derived depend on the assumed grain composition. For a generic grain of refraction index $n$, the scattered field depends on the product $ xn=2\pi a n/\lambda$: an increase of the refraction index $n$ compensated by a similar decrease of the grain radius $a$ provides the same result. Therefore, the uncertainties on the assumed refraction index affect only the absolute value of the grain size, not their ratios.  The evidence for a spatial variation  of the grain sizes at the disk edge appears therefore strong. 

%\begin{deluxetable}{cccc}
%%\tabletypesize{\scriptsize}
%\tablecaption{Dust grain radius at different positions\label{Tab:grainsize}}
%%\tablewidth{20pt}
%\tablehead{ 
%\colhead{Pixel} & \colhead{RA (2000.0)} & \colhead{Dec (2000.0)} & \colhead{a ($\mu$m)} 
%}
%\startdata
%A &5:35:11.349& -5:24:25.55&	    0.60 \\
%B &5:35:11.355& -5:24:25.40&	    0.53 \\
%C &5:35:11.362& -5:24:25.30&	    0.51 \\
%D &5:35:11.365& -5:24:25.25&	    0.49 \\
%E & 5:35:11.369& -5:24:25.20&	   0.40 \\
%F & 5:35:11.372& -5:24:25.15&	   0.20 \\
%\enddata
%% Text for table notes should follow after the \enddata but before
%% the \end{deluxetable}. Make sure there is at least one \tablenotemark
%% in the table for each \tablenotetext.
%\end{deluxetable}

%The fitting procedure was performed by normalizing $r_{obs}$ dividing by the geometrical cross section $G$ obtained from the fitting radius, $Q_{obs} =ln( r_{obs}) / G$. This brings to more regular, easy to handle curves. 

{Our analysis, based on Eq. (2), does not account for a scattering term. Heuristically, we not expect
scattering to be dominant. One the one hand, it is reasonable to assume that our dust particles are non-spherical, randomly oriented, heterogeneous and with variable refraction index. In this case, the oscillations of the individual cross sections predicted by the Mie theory are smeared out and no large-scale chromatic patterns should be expected. Also, scattering by typical disk  grains is highly non-isotropic, as Mie theory predicts 90\% of the scattered light is emitted in the forward hemisphere. Since our source is mostly back-illuminated, wave propagation toward us is dominated by the coherent superposition of the background light with the forward scattered wave, i.e. extinction. Third, even assuming isotropic scattering, it may be shown that the color effect induced by the scattered radiation produces an apparent modulation of the reddening law (W. Henney, private communication) that seems unable to explain the  observed range of attenuation, and therefore of grain size, shown by our data. 
}

\section{Discussion}\label{sec:Discussion}

\subsection{Grain size variation in the photoevaporative wind}
Some evidence for the spatial variation of the size of dust grains has
been obtained by recent interferometric observations of young disks in
the sub-millimeter \citep{Banzatti+11, Guilloteau+11}. The
general result found so far is that larger grains are found in the
disk regions which are closer to the central star, and this is the
same trend found for the 114-426 disk. In those cases, however, the
dust probed by those observations is located inside the disk, and not
in a photoevaporative wind as for the translucent north side of
114-426.

The result presented in this paper is in general agreement with the
expectations for the dynamics of dust grains entrained in a gaseous
photoevaporative wind. The drag force experienced by a dust grain is
proportional to $a^2$ (Takeuchi et al. 2005), and therefore the
acceleration scales with $a^{-1}$. This means that the wind
selectively entrains grains with different sizes, with smaller grains
being more easily dragged along the wind. As shown by \citet{Owen}, 
the drag force on small grains can dominate over the stellar
gravity allowing them to be carried to very large distances from the
central star (several hundreds of AU). This can explain, at least in a
qualitative way, the radial gradient of the grain sizes obtained by
our analysis over a spatial scale of $200-300$~AU (Fig.~\ref{fig:mappa}).

However, the Owen et al. models treat the case of a photoevaporative
wind driven by the ionizing radiation from the central star, whereas in the
case of 114-426 the observed asymmetries indicate that the origin of
the photoevaporative wind is external to the system, i.e. due to the
diffuse FUV radiation field of the nebula (Section 3). This has an
obvious effect on the geometry predicted for the photoevaporation of
the disk. A model which calculates the photoevaporative wind induced
by a diffuse non-ionizing radiation field is needed to obtain a more
quantitative comparison with the data.

\subsection{A scenario for 114-426}

To build a possible scenario for 114-426, we begin by observing that the disk surface appears shielded by the EUV flux from $\theta^1$Ori-C, unlike the nearby proplyds 117-421 and 121-434 (Fig.~\ref{fig5}). This can be explained in two ways. On the one hand,  114-426 may lie much further away from the Trapezium than the other two proplyds, possibly within the foreground Veil of the Orion Nebula, shielded by the EUV photons from $\theta^1$Ori-C by some extended ionization front, possibly at the inward-facing side of the Veil. In this case, it seems difficult to explain the existence of the bright filamentary structure encircling our source on the West side (Fig.~\ref{fig5}), which is basically the direction opposite to $\theta^1$Ori-C.  This  filament, mostly visible in H$\alpha$, most likely represents a ionization front produced by diffuse EUV radiation coming from the outer regions of the nebula, absorbed by a density enhancement apparently centered on our target.  At the same time, there is no evidence of a comparably strong ionization front on the East side, facing $\theta^1$Ori-C. In order to be viable, the "within the Veil" scenario seems to require a rather peculiar combination of geometry, veil density and radiation field.
  
\begin{figure*}
\epsscale{1.0}
\plotone{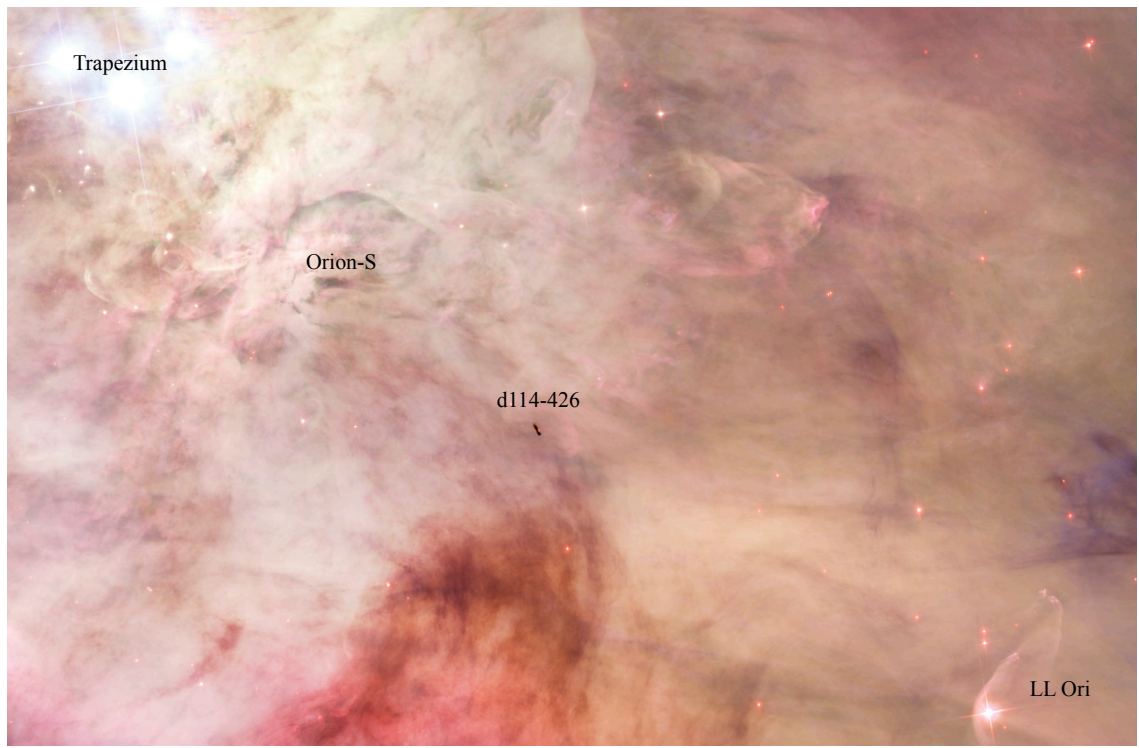}
\caption{Position of 114-426 with respect to the Trapezium, Orion-S and LL Ori}
\label{fig:RegionBig}
\end{figure*}

{The second possibility is that the disk lies in the outer parts of the nebula in a location shielded by the EUV radiation directly coming from $\theta^1$Ori-C.}
It is intriguing, in this respect, to notice that between 114-426 and the Trapezium lies, projected on the plane of the sky, the high-density region hosting Orion-S, a secondary center of star formation in the Orion Nebula (Fig.~\ref{fig:RegionBig}). According to most recent 3-D model of the Orion Nebula \citep{ODell+09,ODellHarris10}, the Orion-S feature southwest of the Trapezium is an optically thick molecular cloud within the body of ionized gas. The Orion-S Cloud that may therefore prevent EUV photons from the Trapezium stars to reach 114-426. The two proplyds 121-434 and 117-421 would therefore lie in the foreground, far enough from the Orion-S Cloud to see directly the Trapezium stars. Alternatively, they could be located deeper inside the Nebula, exposed to the EUV flux passing ''under'' the Orion-S Cloud.
%The shadow effect caused by Orion-S may also explain why the disk side most directly exposed to the diffuse radiation field is the western one. 
%The dark tails seen on the eastern, downwind side appear prominent because of the higher optical depth caused by a somewhat lower outflow velocity;  a dark dust tail in the downwind direction is often observed also in prominent photoevaporated proplyds. 
%Like the region to the West of the Orion-S r is exposed to the diffuse nebular radiation, our disk photoevaporates 

We propose that the influence of the FUV photons mostly coming from the Western regions of the Orion Nebula causes 114-426 to photoevaporate. % and to the plasma flowing away from the main ionization front at the interface with the underlying OMC-1 molecular cloud,  which decreases its  electron density as it accelerates and recombines \citep[see e.g.][]{O'Dell+08}.  
The contribution from other stellar sources to the East of 114-426 is neglible, in particular we estimate that the total flux from closest A0 star KS~Ori (Par 1685) at $\simeq 3$~arcmin distance  is about 1\% of the diffuse FUV field (see the next section).   The neutral outflow from the disk surface intercepts the weak diffuse field of EUV photons at relatively large distances from the disk; still, the ionized filamentary structure may well represent the analog of the bright ionized cusps that identify common proplyds. The pressure gradient pushes the neutral material further to the East, originating``foot-like'' structure traced by a weak ionization front at its boundaries. Also the photoionized proplyds in the vicinity of $theta^1$Ori-C show dusty tails downstream of the ionizing radiation. In summary, 114-426 can be regarded as an extreme case of proplyd photoevaporated by the diffuse radiation field.

%{\bf  Qui alcune cose sono state antiipate nella intro, da rivedere}
%(The following from \emph{Externally illuminated young stellar objects in the Orion Nebula (Bally)}: check!!)

\subsection{Mass loss from 114-426}

The mass loss rate, $\dot{M}$, for large disks that evaporate in an external FUV field is given by \citep{Adams+04}:
\begin{equation}
\dot{M} = 4\pi \mathcal{F}\langle \mu \rangle \sigma^{-1}_{FUV} c_s r_d .
\end{equation} 
where $\mathcal{F}$ is the fraction of solid angle subtended by the outflow, $\mu$ is the mean molecular mass of the gas, $\sigma_{FUV}$ is the absorption cross section for dust grains to FUV radiation,  $c_s$ is the sound speed of the flow, and $r_d$ is the disk radius. With \citet{Adams+04}, we assume $\mathcal{F} \sim 1$. Assuming a typical dust grain radius $a=0.5~\mu$m, as derived in Section~\ref{sec:Dust}, and a gas-to-dust mass ratio of $100:1$ we derive $\sigma_{FUV} = 8.4\times10^{-22}$ cm$^2$, in line with the typical value for PDR regions \citep{StorzerHollenbach99}. For the outer radius we use $r_d \simeq 400$ AU, concentrating our analysis on the northern translucent region. The sound speed, i.e. the temperature at the disk surface, is controlled by the amount of FUV illuminating the disk which can be estimated as follows.

Both ioinizing (EUV) and non-ionizing (FUV) fluxes emitted by the Trapezium stars are scattered within the HII region. The EUV radiation, carrying about 2/3 of the original ionizing energy (assuming Menzell's case B recombination), is scattered by recombination processes, whereas FUV radiation is scattered only by dust particles \citep[see e.g.][]{Robberto+02}. Most of the scattering occurs at the edges of the HII region so that the diffuse field can be assumed to be constant within the cavity. 
A disk shielded from the EUV radiation and from direct illumination from the cluster center will only receive 
a flux of scattered FUV photons given by
\begin{equation}
F_{diff}\sim \frac{a}{2}Q_{sca}\cdot \frac{L_{FUV}}{4\pi R_{HII}^2},
\end{equation} 
assuming that 1/2 of the radiation is redirected inward by the walls of the cavity, a scattering efficiency $Q_{sca}\simeq 1$ appropriate for large grains, a flux $F_{FUV}$ of non-ionizing photons emitted by the central O stars (mostly $\theta^1$Ori-C), and a typical radius of the cavity $R_{HII}=0.3$~pc, the distance between $\theta^1$Ori-C and the ionization front \citep{ODell+08}. 
We assume\footnote{$\theta^1$Ori-C is historically flagged with a spectral type O6.5pe. However, recent observations indicate that this peculiar system is a binary with a O5.5 primary (M=34.0$M_\odot$) and an O9.5 secondary \citep[M=15.5 M$\odot$][]{Kraus+07}. We therefore adopt the Lyman continuum luminosity of a ZAMS O5.5 star, according to \citet{Panagia73}. Our results to not depend on this assumption.} 
that $\theta^1$Ori-C emits $50\%$ of its luminosity as FUV photons,  $L_{EUV} \simeq 1.5\times10^{49}$photons s$^{-1}$ and an effective wavelength $\lambda_{FUV}=1200$~\AA.  A shielded disk will then receive a diffuse FUV flux  $F_{diff}\sim 2 \cdot 10^3 G_0$. 

Solving the equilibrium equation for the radiative heating and cooling of dust grains \citep[e.g.][Ch24]{Draine}, assuming the above FUV flux and the refraction index of our heterogeneous grain, we derive a dust temperature of 25~K, 
appropriate for a Photo Dissociation Region. For the temperature of the gas components, we can refer to Fig.2(b)
of \citet{Adams+04}, which reports the temperature profiles calculated from the PDR code for an external FUV radiation field of $3000~G_0$ as a function of visual extinction $A_V$, for different values of the number density. At $A_V\simeq1$ the gas temperature is in the range $T_g\simeq100-300$\,K, corresponding to a sound speed $c_s \simeq 0.6 - 1.0 $km s$^{-1}$  from the disk. 

Using Equation~(3) we then derive for the mass loss rate of 114-426 
\begin{equation}
\dot{M}  \sim 3-5 \times 10^{-7} \frac{M_{\odot}}{yr}
\end{equation} 
where the spread is mainly dependent on the uncertainty on the sound speed. %a rather high value, considering that our estimate, roughly independent of location within the nebula, may be enhanced by the contribution from the diffuse EUV flux. 

\subsection{Disk lifetime and unicity}

The large disk size, combined with the modest upper limit for the disk mass, make 114-426 a rather unique object. Analyzing the thermal continuum (880~$\mu$m) emission from circumstellar disks in the $\sim1$~Myr old Ophiucus star-forming region, \citet{Andrews+10} have found a strong correlation between the disk mass and radius, larger disks being more massive.  A comparison of our values for the disk radius and the mass upper limit with the valuer reported by \citet{Andrews+10} shows that our source does not follow the radius-mass relation traced by the other sources, being way too large for its mass.  If the relation radius-mass is an imprint of the initial conditions inherited by the disk at the formation epoch, as suggested by \cite{Andrews+10}, then 114-426 is in a diverse evolutionary stage in which this dependence seems to be lost.
%\begin{figure*}
%\epsscale{1.0}
%\plotone{immagini_definitive/Andrews2010fig10.pdf}
%\caption{Position of 114-426 in the plot of \cite{Andrews+10}, with the proposed mass-radius relation for $\sim1$~Myr %old circumstellar disks in Ophiucus (solid line).}
%\label{fig:Andrews}
%\end{figure*}

Assuming our range of estimated mass loss rates and the current upper limit for the disk mass, $1.2\times10^{-2}$M$_\sun$, the disk's life time ranges between $2.2 \times 10^4$~yr to $3.7\times 10^4$~yr. This is clearly a very short time scale, but it may be appropriate for a rather unique object like 114-426. At the current evaporation rate, 114-426 will be entirely dissipated in a few  $\sim 10^4$~yr. It may well have formed 1~Myr ago (about the age of the ONC) with an average disk mass of $\sim 0.1-0.5$~M$_\odot$ and be reduced to the current mass maintaining level of photoevaporation nearly constant at the measured level. Overall, the picture seems consistent with 114-426 being a relatively low mass disk inflated and photoevaporated by the diffuse field of FUV radiation within the nebula, seen in the very last phases of its evolution. 
% in a star forming region of $\sim 10^6$~yr may be observed if there are hundreds of potential candidates, as it seems to be our case. On the other hand, 114-426  

\subsection{Disk warping in the ONC}
In this last section we briefly discuss the disk morphology. First, we note that the keplerian orbital period at the outer disk edge  
\begin{equation}
P_{out}=\sqrt{4\pi^2r_d^3\over{GM_\star}}\simeq 8000{\rm yr}
%print,SQRT(4*!PI^2*(400.*1.5D13)^3/(6.67E-8*2E33))/3.15E7
 %      8026.3539
\end{equation}
corresponds, for dust grains moving at 1km/s, to a traveled distance of about 1700~AU, or 4 disk radii. 
%IDL> print,SQRT(4*!PI^2*(400.*1.5D13)^3/(6.67E-8*2E33))*1E5/1.5E13
 %      1685.5343
This is nicely compatible with the spatial scale of variation of the spiraling dust filaments on the eastern side of the disk, which may therefore be related to periodic enhancement of the disk mass loss rate syncronous with the passage, or appearance, of disk asymmetries.

Concerning the disk asymmetry,while the diffuse FUV radiation represents the main mechanism of destruction of 114-426,  it cannot easily account for the extremely warped disk geometry unless one assumes it is highly non-isotropic. A  directional radiation flux impacting the surface of a flared disk can generate a non-uniform increase of the surface temperature and modulate the flaring angle \citet{Robberto+02}, or just push the disk mid-plane out of its equilibrium position, depending on the disk angular velocity and surface density (S. Lubow, private communication). Since the disk is rotating, the development of a persistent warp will depend on factors like the cooling time scale, the turbulence regime, or any possible temporal variations of the radiation intensity, all basically unknown at the moment.

A second possibility is that the disk hosts a binary star, in which case the warping would result from a a misalignement between the disk plane and the plane of the binary orbit. Very little is known about the nature of the central star \citep[see][for a discussion]{McCaughrean98}. An intriguing aspect is that the colors of the two lobes are quite different: the south-eastern lobe is redder, to the point of becoming prominent in the near-IR, as revealed by the HST/NICMOS images of \citet{McCaughrean98}. It is not possible to explain such an inversion by assuming a standard flared disk model and common optical properties (scattering/absorption efficiencies) for the dust grains. However, a binary system with two stars having comparable luminosity and different effective temperatures could represent a more viable solution. Direct spectroscopy of the two lobes can address this possibility. 

Stellar encounters may also perturb circumstellar disks leading to the fomation of warps, spiral waves, tidal tails and disk truncation \citep{Zinnecker+Yorke07}.
%mr: controlla questa referenza, e' una review.  Altre su Nixon+Pringle2010
Those disturbances, observed e.g. in RW Aur by \citet{Cabrit+06}, are expected to occur in the ONC.
%altre ref in Nixon+Pringle Sez.1
Several studies have pointed out that in the dense inner regions of young clusters like the ONC stellar enconters may be frequent enough to produce major effects on circumstellar disks \citep[see e.g.][and reference therein]{Olczak+2008}.

Warps induced in the outer disk regions by the fly-by of another star propagate inward to a radius $r<r_d$ with a wavelength given by
\citep{LubowOgilve+Pringle2002}
\begin{equation}
\lambda_w(r)\sim {c_s(r)\over{2V}D}
\end{equation} 
where $c_s(r)$ is the  local disk sound speed and $V$ and $D$ are the relative velocity and closest approach distance of the encounter, respectively. The warp propagate inward to a radius $r_{crit}$ at which $r=\lambda(r)$, where the wavelength of the warp becomes larger than the disk size and gets reflected. Angular momentum, however, is still communicated to the inner disk so the disk becomes tilted, but not warped. The decay time-scale for the warp is given approximately by 
\begin{equation}
t_{damp}\sim{P_{out}\over{2\pi\alpha}}
\end{equation}
where 
$\alpha$ is the dimensionless viscosity parameter, typically in the range $\alpha\sim10^{-2}-10^{-4}$\citep{Hartmann09}. For our disk with $r_d=400$AU, assuming $M_\star=1$~M$_\sun$, the warp may be expected to last for a time of around $1.3\time10^6(\alpha/10^{-4})$yr. 
%IDL> print,SQRT(4*!PI^2*(400.*1.5D13)^3/(6.67E-8*2E33))/(2*!PI*1E-4)/3.15E7
 %      12774339.
Therefore, for the low values of the viscosity parameter $\alpha$ typical of the outer disk regions, a warp imprinted to the disk by a past close encounter may still be  propagating at these later times. 

An interplay between the warped outer disk, the flat and tilted inner disk, the energy input from the central star(s) and the nebular environment supporting a conspicuous photoevaporative flow may account for the  complex phenomenology observed in this unique system. In the near future, ALMA observations should greatly help clarifying the nature of 114-426 by mapping with high spatial resolution the spatial distribution of the disk and its kinematic status. 
 
%{\bf [Here is some inbput from Steve Lubow... Lodato may add on binarity.... put this discussion in a second paper with
%both of them?]}
%Before concluding, let us discuss briefly the disk warping. 
%Consider a very thin disk subject to an external force such as a wind
%with force per unit area perpendicular to the disk $F_n$. Consider a  
%coordinate system with the unperturbed disk to lie midplane along the the $z=0$ plane.
%The disk midplane can be shown to be pushed from its equilibrium  
%position by an amount 
%$$
%z = \frac{F_n}{\Omega^2 \Sigma}
%$$
%where $\Omega$ is the local disk
%rotation rate, assumed Keplerian, and Sigma is the local disk surface  
%density.
%In the case of a wind, $F_n = \rho v_n^2$, where $v_n$ is the velocity
%perpendicular to the disk and $\rho$ is the wind density near the disk.
%So the disk midplane shift/warp can be translated into a
%ocal measure of external force across the disk surface.

\section{Conclusions}\label{sec:Conclusions}
We have used a set of HST images taken with WFPC2 and ACS to describe the complex morphology of the most prominent dark silhouette disk in the Orion Nebula, 114-426. The disk appears tilted, asymmetric and warped, with clear signs of photoevaporation of neutral material. 
Mapping the distribution of dust grains at the northern translucent edge of the disk using the Mie theory for standard circumstellar disk grains, we have found evidence for a spatial gradient of the grain radius, from 0.2 to 0.7~$\mu$m,
with grains becoming smaller as the distance from the disk increases. These values are slightly below the minimal 1.9~$\mu$m grain size estimated by Shuping et al. (2003), but still compatible considering the difficulty of observing this source at infrared wavelengths.
%Mapping the distribution of dust grains at the northern translucent edge of the disk using the Mie theory for standard circumstellar disk grains, we have found evidence for a spatial gradient in grain radius, \citep[$\simeq 0.2-0.7~\mu$m, less than what found by][]{Shuping}, with grains becoming smaller as the distance from the disk increases. The observed spatial distribution of grain size is consistent with the expectations for the dynamic of dust entrained in a gaseous photoevaporative wind.
Overall, the disk morphology and location in the nebula indicate that this system is photoevaporated by the diffuse radiation field of the Orion Nebula, while being shielded from the radiation coming directly from the central Trapezium stars. Using our derived grain properties, we  have estimated the mass loss rate from the disk surface and the time-scale for total disk dissipation, which turns out to be of the order of $10^4$yr. This short time scale, of the order of 1/100 of the cluster age,  seems compatible with the exceptional nature of this system, namely its huge size and low mass. Finally, we briefly discuss the viability of possible mechanisms that may lead to the peculiar morphology of this system: external UV flux, binary star and past close encounter.

\vskip 0.5cm
{\sl Acknowlegments}: the authors would like to thank S. Lubow, I. Pascucci, J. Pringle and K. Stapelfeld for helpful discussions on this fascinating source, and the referee W. Henney for prompt and useful feedback. Zolt Levay provided support with image production. Support for GO program 10246 was provided by NASA through a grant from the Space Telescope Science Institute, which is operated by the Association of Universities for Research in Astronomy, Inc., under NASA contract NAS 5-26555.

%_____________________________________________________________________________________

\end{document}